\documentclass{emulateapj}
\newcommand{\su}{\textit{Suzaku}}

\usepackage{color}

\slugcomment{Draft Version \today}

\shorttitle{METALLICITY OF TYCHO'S PROGENITOR}
\shortauthors{Badenes, Bravo \& Hughes}

\begin{document}

\title{The End of Amnesia: A New Method for Measuring the Metallicity of Type Ia Supernova Progenitors Using Manganese Lines
  in Supernova Remnants}

\author{Carles Badenes\altaffilmark{1,2}, Eduardo Bravo\altaffilmark{3} and John P. Hughes\altaffilmark{4}}

\altaffiltext{1}{Department of Astrophysical Sciences, Princeton University. Peyton Hall, Ivy Lane, Princeton NJ 08544-1001; badenes@astro.princeton.edu}

\altaffiltext{2}{\textit{Chandra} Fellow}

\altaffiltext{3}{Departament de F\'{i}sica i Enginyeria Nuclear, Universitat Polit\`{e}cnica de Catalunya, Diagonal 647,
  Barcelona 08028, Spain; and Institut d'Estudis Espacials de Catalunya, Campus UAB, Facultat de Ci\`{e}ncies,
  Bellaterra, Barcelona 08193, Spain; eduardo.bravo@upc.es}

\altaffiltext{4}{Department of Physics and Astronomy, Rutgers University. 136 Frelinghuysen Rd., Piscataway NJ 08854-8019; jph@physics.rutgers.edu}

\begin{abstract}
  We propose a new method to measure the metallicity of Type Ia supernova progenitors using Mn and Cr lines in the X-ray
  spectra of young supernova remnants.  We show that the Mn to Cr mass ratio in Type Ia supernova ejecta is tightly
  correlated with the initial metallicity of the progenitor, as determined by the neutron excess of the white dwarf
  material before thermonuclear runaway. We use this correlation, together with the flux of the Cr and Mn K$\alpha$
  X-ray lines in the Tycho supernova remnant recently detected by \su\ \citep{tamagawa08:Tycho_Suzaku} to derive a
  metallicity of $log(Z)=-1.32^{+0.67}_{-0.33}$ for the progenitor of this supernova, which corresponds to
  $log(Z/Z_{\odot})=0.60^{+0.31}_{-0.60}$ according to the latest determination of the solar metallicity by
  \citet{asplund05:solar_composition}. The uncertainty in the measurement is large, but metallicities much smaller than
  the solar value can be confidently discarded. We discuss the implications of this result for future research on Type
  Ia supernova progenitors.
\end{abstract}


\section{INTRODUCTION}
\label{sec:Intro}

Despite decades of continuing effort, the nature of the progenitor systems of Type Ia supernovae (SNe) remains
unknown. The most widely accepted theoretical scenarios involve the thermonuclear explosion of a C+O white dwarf (WD)
destabilized by accretion of material from a binary companion, either another WD (double degenerate systems, DDs) or a
normal star (single degenerate systems, SDs). Observations of supernova rates in distant galaxies suggest at least two
different ways to produce Type Ia SNe \citep{scannapieco05:A+B_models}: a `prompt' channel associated with young stellar
populations and a `delayed' channel associated with old stellar populations. At present, it is not clear what
relationship, if any, these channels have with the theoretical scenarios, or even if the proposed scenarios can yield
Type Ia SNe at the observed rate \citep{maoz08:fraction_intermediate_stars_Ia_progenitors}. Most attempts to constrain
the fundamental properties of the progenitors have been unsuccessful, including direct identification in optical
pre-explosion images \citep{maoz08:Ia_progenitor_search_NGC1316}, searches for H stripped from the binary companion in
SN spectra \citep{leonard07:H_nebular_Ia_spectra}, and searches for prompt emission from circumstellar interaction
\citep{panagia06:radio_SNIa,hughes07:X_Ray_Flux_SNIa}. Very recently, the possible identification of the progenitor in a
pre-explosion X-ray image of SN 2007on by \citet{voss08:SN2007on_progenitor} has been questioned by post-explosion
images \citep{roelofs08;no_SN2007on_progenitor}. Campaigns to look for the surviving companion star to the exploded WD
in the nearby Tycho supernova remnant (SNR), known to be of Type Ia origin, have also produced controversial results
\citep{ruiz-lapuente04:Tycho_Binary,ihara07:Tycho_Companion}. This lack of clues to the identity of the progenitors in
the observations of Type Ia SNe is sometimes referred to as `stellar amnesia'.

One of the most important constraints on the age and nature of the progenitor systems is their metallicity $Z$. During
stellar evolution, the C, N, and O that act as catalysts for the CNO cycle pile up into $^{14}$N, which is converted to
$^{22}$Ne in the hydrostatic He burning phase through the reactions
$^{14}$N($\alpha$,$\gamma$)$^{18}$F($\beta^+$,$\nu_\mathrm{e}$)$^{18}$O($\alpha$,$\gamma$)$^{22}$Ne. The $\beta^+$ decay
of $^{18}$F increases the neutron excess of the WD material, defined as $\eta=1-2 \langle Z_{A}/A \rangle$ (where
$Z_{A}$ is the atomic number and $A$ is the mass number), resulting in a linear scaling of $\eta$ with metallicity:
$\eta=0.101 \times Z$ \citep{timmes03:variations_peak_luminosity_SNIa}. This can have an observable impact on Type Ia
supernova spectra \citep{lentz00:metallicity_effects_SNIa}, but the predicted effects are hard to detect in practice
\citep[in one case, SN 2005bl, a progenitor of subsolar metallicity has been suggested
by][]{taubenberger08:SN2005bl}. In this \textit{Letter}, we describe a new direct method to measure the metallicity of
Type Ia SN progenitors using Mn and Cr lines from shocked ejecta in the X-ray spectra of SNRs.

\section{THE $M_{Mn}/M_{Cr}$ RATIO AS A TRACER OF PROGENITOR METALLICITY}
\label{sec:metallicity}

\subsection{Mn/Cr vs. $Z$}

Nuclear burning in Type Ia SNe happens in different regimes depending on the peak temperature reached by the WD
material. From higher to lower temperatures, the regimes are: nuclear statistical equilibrium (NSE), incomplete Si
burning, incomplete O burning and incomplete C-Ne burning \citep{woosley86:SAAS-FEE}. During the explosion itself,
electron captures are too slow to change the original value of $\eta$ except in the innermost
$\sim0.2\,\mathrm{M_{\odot}}$ of ejecta, where nuclear burning happens at higher densities, leading to neutron-rich NSE
\citep{brachwitz00:electron_captures_SNIa}. Outside this region, an imbalance between free neutrons and protons does not
affect the production of the most abundant species, with the possible exception of $^{56}$Ni at supersolar metallicities
\citep{timmes03:variations_peak_luminosity_SNIa}. However, some trace nuclei with unequal numbers of protons and
neutrons are very efficient at storing the neutron excess. The most abundant among these nuclei is $^{55}$Mn, which is
produced during incomplete Si burning as $^{55}$Co. When normalized to another product of incomplete Si burning whose
nucleosynthesis is insensitive to the neutron excess (Cr is the ideal choice), the yield of Mn becomes an excellent
tracer of the progenitor metallicity.

\begin{figure}

  \centering

  \includegraphics[angle=90,scale=0.9]{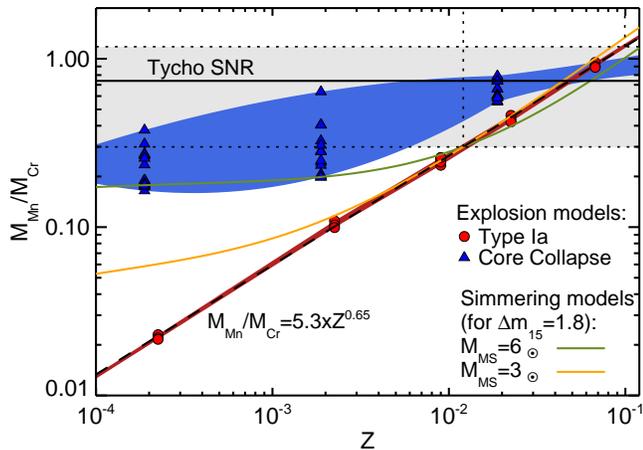}  

  \caption{Mn to Cr mass ratio as a function of progenitor metallicity for Type Ia \citep[][this
    work]{badenes03:xray,badenes05:xray} and core collapse \citep{woosley95:core_collapse_models} SN models, represented
    with red circles and blue triangles, respectively. The red and blue shaded regions encompass all the models in each
    class. The dashed line represents a power-law fit to the Type Ia models. The orange and green plots illustrate the
    impact of weak interactions during the simmering phase for very subluminous Type Ia SNe ($\Delta m_{15}=1.8$) in WDs
    with main sequence progenitor masses of $3\,\mathrm{M_{\odot}}$ and $6\,\mathrm{M_{\odot}}$, respectively (see
    \S~\ref{sub:simmering}). The horizontal solid and dotted lines represent the estimated $M_{Mn}/M_{Cr}$ ratio in the
    Tycho SNR from the \su\ observations ($0.74\pm0.47$, shaded in gray), which corresponds to
    $log(Z)=-1.32^{+0.67}_{-0.33}$ (see \S~\ref{sec:ZofTycho}). \label{fig-1}}
  
  \vspace{0.25cm}

\end{figure}

This diagnostic can be quantified with Type Ia SN explosion models. We have used the model grid from
\citet{badenes03:xray,badenes05:xray}, complemented with other models calculated with the code described in
\citet{bravo96:code}. Physical inputs and initial conditions are as in \citet{badenes03:xray}, except in the cases noted
below. We consider examples of several explosion mechanisms: deflagrations, delayed detonations, and pulsating delayed
detonations. For the present work, we have recalculated the nucleosynthesis of four delayed detonation models (DDTa,
DDTc, DDTe, and DDTf) at different metallicities by altering the value of $X(\mathrm{^{22}Ne})$ in the pre-explosion WD
from the canonical $0.01$ (corresponding to $Z=9.0 \times 10^{-3}$). We have also calculated one deflagration model
starting from a central density twice as large as the other models, and one delayed detonation model with the C to O
mass ratio $M_{C}/M_{O}=1/3$ in the WD instead of the usual 1. In spite of these fundamental differences in explosion
mechanisms and initial conditions, we find a very tight correlation between the Mn to Cr mass ratio outside the
neutron-rich NSE region in the SN ejecta, $M_{Mn}/M_{Cr}$, and the progenitor metallicity $Z$ (see Figure
\ref{fig-1}). A power-law fit yields the following relation (with a correlation coefficient $r^{2}=0.9975$):

\begin{equation}
M_{Mn}/M_{Cr}=5.3 \times Z^{0.65} 
\label{eq-1}
\end{equation}

Removal of the inner $0.2\,\mathrm{M_{\odot}}$ of neutron-rich NSE material from this relation is justified because (1)
there is observational evidence that this material does not mix outwards during the explosion
\citep{gerardy07:SNIa_midIR,mazzali07:zorro} or its aftermath \citep{fesen07:SN1885}; and (2) the dynamical ages of most
ejecta-dominated Type Ia SNRs are too small for the reverse shock to have reached so deep into the ejecta. For
illustration, we have plotted in Figure \ref{fig-1} the core collapse SN models from
\citet{woosley95:core_collapse_models}, which also show some correlation between $M_{Mn}/M_{Cr}$ and $Z$, albeit with a
much larger spread and a much shallower slope.

\subsection{The impact of C simmering}
\label{sub:simmering}

In order for Eq. \ref{eq-1} to hold, the value of $\eta$ set by the progenitor metallicity must remain unchanged between
the formation of the WD and the SN explosion. This should be true for DD progenitors if the final merger and runaway
happen on dynamical timescales. In slowly accreting WDs, $\eta$ can be modified through electron captures during the
so-called `simmering' phase of non-explosive C burning that takes place in the $\sim 1000$ yr prior to the explosion
\citep{piro08:neutronization_SNIasimmering}. The impact of this additional neutronization on the $M_{Mn}/M_{Cr}$ ratio
will depend on the extent of the convective region over which the neutronized material is mixed ($M_{conv}$), and its
overlap with the portion of the WD that will undergo incomplete Si burning where Mn and Cr are synthesized.

\begin{figure}

  \centering

  \includegraphics[angle=90,scale=0.9]{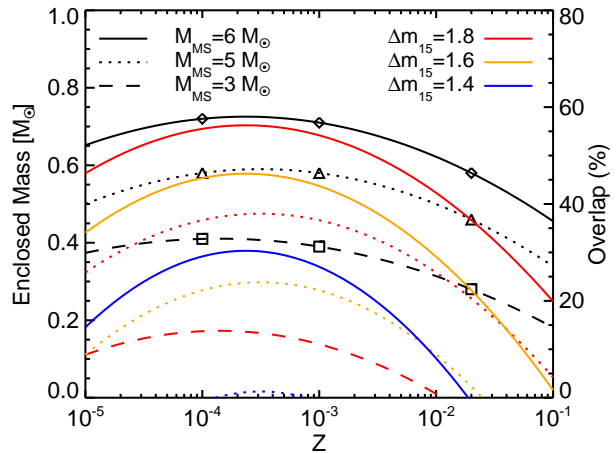}  

  \caption{Upper limits to $M_{conv}$ during the simmering phase (left axis, black plots) and its overlap with the Si
    rich region of ejecta (right axis, colored plots) as a function of $Z$. The black plots represent quadratic
    interpolations for different $M_{MS}$: $3\,\mathrm{M_{\odot}}$ (dashed), $5\,\mathrm{M_{\odot}}$ (dotted), and
    $6\,\mathrm{M_{\odot}}$ (solid), with the symbols indicating the values from \citet{dominguez01:SNIa_progenitors}
    (for $M_{MS}=6\,\mathrm{M_{\odot}}$ at $Z=0.02$, we have taken the average between $M_{MS}=7\,\mathrm{M_{\odot}}$
    and $M_{MS}=5\,\mathrm{M_{\odot}}$). The colored plots represent the overlap (in $\%$) between $M_{conv}$ and the
    Si-rich region of ejecta, for very subluminous ($\Delta m_{15}=1.8$, red), mildly subluminous ($\Delta m_{15}=1.6$,
    orange), and normal, but faint ($\Delta m_{15}=1.4$, blue) Type Ia SNe. \label{fig-2}}

\end{figure}

Some previous studies
\citep{kuhlen06:C_ignition_SNIa,piro08:neutronization_SNIasimmering,chamulak08:reduction_electron_simmering_SNIa} have
found large values for $M_{conv}$ by assuming an homogeneous chemical composition for the WD, effectively applying the
Schwarzschild criterion for convection. We will adopt the hypothesis that the convective region is limited by the Ledoux
criterion to the C-depleted core of the WD created during hydrostatic He-shell burning, $M_{core}$
\citep{hoeflich02:runaway}. In the absence of a self-consistent picture for convection inside WDs, this must remain an
open issue \citep[see][for a discussion]{piro08:convective_core_radius}, but we believe that our hypothesis is
reasonable given the existence of other processes that can reduce the extent of $M_{conv}$ and limit the mixing of
neutronized material. These include the presence of Urca shells \citep{stein06:convective_urca} and the long thermal
diffusion time scale for buoyant bubbles larger than $\sim 100$ m \citep{garcia-senz95:SNIa_flame_is_born}, which
prevents them from mixing completely with their surroundings during convection.

\begin{figure}

  \centering

  \includegraphics[angle=0,scale=0.4]{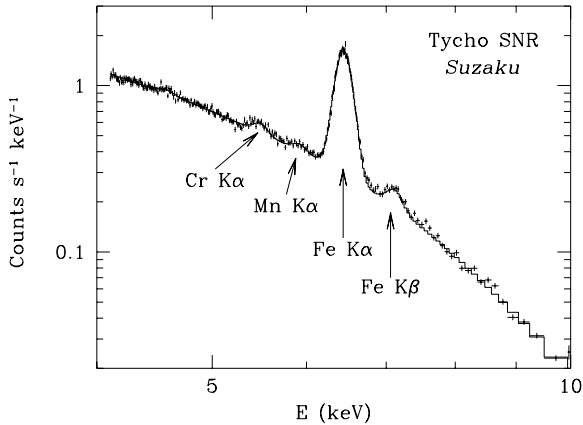}  

  \caption{X-ray spectrum of the Tycho SNR observed by \su\ in the vicinity of the Fe K$\alpha$ line. The K$\alpha$
    lines of Cr and Mn are detected at $>10 \sigma$ and $7 \sigma$ confidence levels, with fluxes of
    $2.45^{+0.48}_{-0.42} \times 10^{-5}$ and $1.13^{+0.38}_{-0.45} \times 10^{-5}\,\mathrm{photons\,cm^{-2}\,s^{-1}}$
    respectively \citep{tamagawa08:Tycho_Suzaku}. \label{fig-3}}

\end{figure}

According to \citet{mazzali07:zorro}, Si-rich ejecta in Type Ia SNe lie between Lagrangian mass coordinates
$1.05\,\mathrm{M_{\odot}}$ and $(1.55-0.69 \times \Delta m_{15}) \,\mathrm{M_{\odot}}$, where $\Delta m_{15}$ is the
light curve width parameter (defined as the decline in blue magnitude 15 days after maximum), which acts as a proxy for
SN brightness. In this context, the extent of the overlap between $M_{conv}$ and the Si-rich ejecta depends on the main
sequence mass $M_{MS}$ and initial metallicity of the WD progenitor \citep[which determine
$M_{core}$,][]{dominguez01:SNIa_progenitors} and the SN brightness (see Figure \ref{fig-2}). This overlap is only
significant for subluminous ($\Delta m_{15} \geq 1.6$) Type Ia SNe originated by progenitors with either large $M_{MS}$
or low $Z$, or both. \citet{chamulak08:reduction_electron_simmering_SNIa} find an upper limit for the increase of $\eta$
during the simmering phase of $\Delta \eta=0.0015$, which is comparable to the value of $\eta$ in solar material. The
impact of simmering on the $M_{Mn}/M_{Cr}$ ratio can then be estimated by mixing material with $M_{Mn}/M_{Cr}=0.3$
\citep[appropriate for the value of $Z_{\odot}$ derived by][]{asplund05:solar_composition} into the incomplete Si
burning region, in a proportion equivalent to the extent of the overlap shown in Figure \ref{fig-2}. The green and
orange plots in Figure \ref{fig-1} are two examples of such `simmering-modified' models for very subluminous ($\Delta
m_{15}=1.8$) Type Ia SNe, illustrating our conclusion that C simmering will only modify the $M_{Mn}/M_{Cr}$ ratio in the
SN ejecta for very subluminous SNe, and then only in cases where $M_{MS}$ is large, or $Z$ is low, or both.
%In any case,
%it is worth noting that values of $M_{Mn}/M_{Cr}$ below $0.1$ would offer a direct test of the extent of $M_{conv}$ in a
%simmering WD.

\section{MEASURING THE $M_{Mn}/M_{Cr}$ RATIO IN SNRs}

The work in this \textit{Letter} is motivated by the recent \su\ detection of Mn and Cr in the X-ray spectrum of the
Tycho SN reported by \citet{tamagawa08:Tycho_Suzaku} (see Figure \ref{fig-3}). This observational result opens the possibility of
studying the $M_{Mn}/M_{Cr}$ ratio in Type Ia SN ejecta, which cannot be done using optical SN spectra due to the 2.7 yr
half-life of $^{55}$Fe in the decay chain $^{55}$Co$\rightarrow ^{55}$Fe$\rightarrow ^{55}$Mn. Since Mn and Cr are
synthesized together in the explosion and have very similar electronic structures, it is possible to estimate their mass
ratio from the line flux ratio: $M_{Mn}/M_{Cr}=1.057 \times (F_{Mn}/F_{Cr})/(E_{Mn}/E_{Cr})$, where $1.057$ is the ratio
of atomic masses, $F_{Mn}/F_{Cr}$ is the line flux ratio, and $E_{Mn}/E_{Cr}$ is the ratio of specific emissivities per
ion.

\begin{figure}
  
  \centering
  
  \includegraphics[angle=90,scale=0.9]{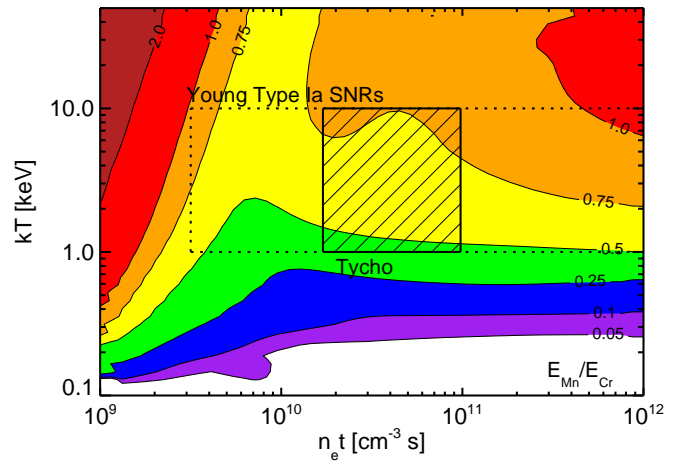}  
  
  \caption{Interpolated Mn to Cr specific K$\alpha$ emissivity ratio as a function of $n_{e}t$ and $kT$. The dotted box
    encompasses the values of $n_{e}t$ and $kT$ found in the Si-rich ejecta of six young Type Ia SNRs by
    \citet{badenes07:outflows}: $9.49 \leq log(n_{e}t) \leq 11.94 \,\mathrm{cm^{-3}s}$; $1.0 \leq kT \leq 10.0
    \,\mathrm{keV}$. The striped area corresponds to the region of parameter space appropriate for the Tycho SNR, $10.23
    \leq log(n_{e}t) \leq 10.99 \,\mathrm{cm^{-3}s}$; $1.0 \leq kT \leq 10.0 \,\mathrm{keV}$. \label{fig-4}}

\end{figure}

Public data bases for X-ray astronomy do not usually include lines from trace elements like Mn and Cr, but the value of
$E_{Mn}/E_{Cr}$ can be estimated by interpolation along the atomic number sequence from elements with available data
\citep{hwang00:W49B}. We have used the ATOMDB data base \citep{smith01:H_like_and_He_like_ions} to retrieve K$\alpha$
line emissivities for Si, S, Ar, Ca, Fe, and Ni ($Z_{A}=14,\,16,\,18,\,20,\,26,\,28$) as a function of ionization
timescale $n_{e}t$ and electron temperature $kT$, the two variables that control the line emission of a plasma in
nonequilibrium ionization. Then we performed a spline interpolation to obtain $E_{Mn}/E_{Cr}$, which we plot in Figure
\ref{fig-4}, together with the region of the $(n_{e}t,kT)$ parameter space that is populated by young Type Ia SNRs
\citep[see the Appendix in][]{badenes07:outflows}. The uncertainty in the value of $E_{Mn}/E_{Cr}$ comes from the
variation within this region and the error introduced by the interpolation itself, which we have estimated by comparing
the true and interpolated values for the $E_{Ar}/E_{S}$ ratio. For the larger region of the $(n_{e}t,kT)$ plane plotted
in Figure \ref{fig-4}, we find $E_{Mn}/E_{Cr}=0.69\pm0.32$ ( $E_{Mn}/E_{Cr}=0.66\pm0.26$ for the smaller region
appropriate for Tycho).

\section{DISCUSSION AND CONCLUSIONS: THE METALLICITY OF TYCHO'S PROGENITOR}
\label{sec:ZofTycho}

From the \su\ observation, $F_{Mn}/F_{Cr} = 0.46\pm0.21$, which gives $M_{Mn}/M_{Cr}=0.74\pm0.47$. This mass ratio
translates into a metallicity of $Z=0.048^{+0.051}_{-0.036}$ for the progenitor of the Tycho SNR
($log(Z)=-1.32^{+0.67}_{-0.33}$). The large error bar in this result is dominated by the statistical uncertainties in
the X-ray fluxes from the faint Mn and Cr lines, and it will be reduced in an upcoming, approved deeper \su\
observation. At present, we find a strong indication for a supersolar metallicity
($log(Z/Z_{\odot})=0.45^{+0.31}_{-0.60}$ with the solar value from \citet{grevesse98:solar_composition};
$log(Z/Z_{\odot})=0.60^{+0.31}_{-0.60}$ with the newer value from \citet{asplund05:solar_composition}). Our results are
compatible with a solar metallicity, but subsolar values can be discarded with confidence. Given this measurement and
the fact that Tycho's SN was probably either normal or slightly overluminous
\citep{badenes06:tycho,ruiz-lapuente04:TychoSN}, we do not have to concern ourselves with the impact of neutronization
during C simmering in this particular case. The Tycho SNR is $59\,\mathrm{pc}$ above the Galactic plane at a
Galactocentric radius of $9.4\,\mathrm{kpc}$ \citep[assuming a distance of
$2.4\,\mathrm{kpc}$,][]{smith91:six_balmer_snrs}. A supersolar metallicity is higher than average for this location, but
well within the spread of measured $\mathrm{[Fe/H]}$ values \citep{nordstrom04:solar_neighbourhood}. This combination of
Galactrocentric radius and metallicity suggests a young progenitor age (a few Gyr or less), which would make Tycho a
candidate for the `prompt' channel of Type Ia SNe, but the scatter in the age-metallicity relations and the
uncertainties in our measurement are too large to be more specific on this point.

In this \textit{Letter}, we have proposed a new method to measure the metallicity of Type Ia SN progenitors using Mn and
Cr lines in the X-ray spectra of their SNRs. We have applied it to the Tycho SNR and obtained a strong indication that
its progenitor had a solar or supersolar metallicity. The main strength of our method is its simplicity: it is based on
well-known nuclear physics and observable parameters that are easy to measure. Detection of Mn and Cr lines in other
young Type Ia SNRs in the Galaxy should be possible with \su, \textit{Chandra}, and \textit{XMM-Newton} observations. A
few additional SNRs in the Magellanic Clouds might be reached with deep exposures, bringing the total of potential
targets to perhaps a dozen. This number would increase significantly with the inclusion of SNRs in nearby galaxies like
M31 and M33, which may be accessible to next generation X-ray observatories like \textit{Constellation-X} and
\textit{XEUS}. Even with a small sample of objects, we should be able to use this method to verify and complement the
indirect metallicity studies of extragalactic Type Ia SNe
\citep{gallagher05:chemistry_SFR_SNIa_hosts,prieto08:SN_Progenitors_Metallicities} and test claims about the metallicity
dependence of the Type Ia SN rate \citep{kobayashi98:lowZ_inhibition_IaSNe}. Measurements or upper limits below a
certain threshold ($M_{Mn}/M_{Cr} \lesssim 0.1$) would also provide interesting constraints on the extent of the
convective region in accreting WDs. We conclude by noting that Mn and Cr lines have already been detected by
\textit{ASCA} in the Galactic SNR W49B \citep{hwang00:W49B}. The measured line flux ratio also suggests a solar or
supersolar progenitor metallicity, but both the age and the SN type of W49B are controversial
\citep{hwang00:W49B,badenes07:outflows}. We defer a detailed discussion of this object and the application of our method
to other Galactic SNRs to a forthcoming publication.

\acknowledgements{The authors are grateful to Inese Ivans, Bruce Draine, and Jim Stone for discussions. We are also
  happy to acknowledge the work of the \su\ Science Working Group and the members of the \su\ Tycho team. Support for
  this work was provided by NASA through Chandra Postdoctoral Fellowship Award Number PF6-70046 issued by the Chandra
  X-ray Observatory Center, which is operated by the Smithsonian Astrophysical Observatory for and on behalf of NASA
  under contract NAS8-03060. EB is supported by grants AYA2007-66256 and AYA2005-08013-C03-01. JPH is partially
  supported by NASA grant NNG05GP87G.}

\bibliographystyle{apj}

\end{document}